# Probing Interface of Perovskite Oxide Using Surface-specific Terahertz Spectroscopy

Yudan Su,[1,2,†] Jiaming Le,[1,†] Junying Ma,[1,†] Long Cheng,[3] Yuxuan Wei,[1] Xiaofang Zhai,[3] Chuanshan Tian[1,*]

[1] Department of Physics, State Key Laboratory of Surface Physics and Key Laboratory of Micro- and Nano-Photonic Structure (MOE), Fudan University, Shanghai 200433, China.

[2] Department of Physics, University of California, Berkeley, California 94720, USA.

[3] School of Physical Science and Technology, ShanghaiTech University, Shanghai 201210, China.

[†] These authors contributed equally to this work.

**Abstract:** The surface/interface species in perovskite oxides play an essential role in many novel emergent physical phenomena and chemical processes. With low eigen-energy in the terahertz region, such species at buried interfaces remain poorly understood due to the lack of feasible experimental techniques. Here, we show that vibrational resonances and two-dimensional electron gas at the interface can be characterized using surface-specific nonlinear spectroscopy in the terahertz range. This technique uses intra-pulse difference frequency mixing (DFM) process, which is allowed only at surface/interface of a medium with inversion symmetry. Sub-monolayer sensitivity can be achieved using the state-of-the-art detection scheme for the terahertz emission from surface/interface. As a demonstration, Drude-like nonlinear response from the two-dimensional electron gas emerging at $LaAlO_3/SrTiO_3$ or $Al_2O_3/SrTiO_3$ interface was successfully observed. Meanwhile, the interfacial vibrational spectrum of the ferroelectric soft mode of $SrTiO_3$ at 2.8 THz was also obtained that was polarized by the surface field in the interfacial region. The corresponding surface/interface potential, which is a key parameter for $SrTiO_3$-based interface superconductivity and photocatalysis, can now be determined optically via quantitative analysis on the polarized phonon spectrum. The interfacial species with resonant frequencies in the THz region revealed by our method provide more insights into the understanding of physical properties of complex oxides.

*Correspondence to: cstian@fudan.edu.cn (C.T.).

**Introduction**

The surface and interface of complex oxides attract enormous research attention due to their unique electrical, magnetic, and electrochemical properties[1,2]. Among these oxides, strontium titanate [$SrTiO_3$ (STO)], as a prototypical perovskite retaining its multifunctional nature and the uniqueness in fabrication and modification with atomic-level precision, stands out as an ideal test-bed for exploring multifarious intriguing physical and chemical phenomena[3], in which the collective excitation and coupling present at surface/interface play the key role[4,5]. For example, at FeSe/STO interface, it is believed that both low-frequency phonon from STO and interfacial band bending are vital for the enhancement of superconductivity[6,7]. The nontrivial topological vortex/antivortex forming at the interface of $PbTiO_3/SrTiO_3$, the collective resonance of which lies in the terahertz (THz) range, provides an alternative choice for post-Moore electronic devices[8]. In the case of STO-based photocatalysis, the facet-dependent surface potential of STO is considered as the essential factor for electron-hole separation and charge transfer across the interface, through which water splitting with almost unity quantum efficiency can be realized[9-11]. However, the interrogation of surfaces/interfaces of perovskite oxides remains challenging experimentally because the fundamental excitations often occurs in the THz region. Yet, little surface-specific probes with chemical selectivity is available below 15 THz, especially, in hostile environment.

Second-order nonlinear optical spectroscopy, such as sum-frequency spectroscopy (SFS), has found multidisciplinary applications in the study of surfaces and interfaces. Being an all-optical detection scheme, it can be used to probe electronic or vibrational resonances in various complex interfacial systems with sub-monolayer sensitivity[12]. Unfortunately, for resonant frequency below 15 THz, employing SFS is extremely difficult because of lacking intense THz light source or a feasible detection scheme that can distinguish the weak sum-frequency signal from the pump light. As a result, over the past decades, applications of surface-specific nonlinear optical spectroscopy were limited to the systems with resonant frequencies ranging from mid-infrared to ultraviolet. On the other hand, there are many important excitations in the THz range,[13, 14] e.g. lattice vibrations, quasi-particles in quantum materials, and hydrogen-bond vibrations in bio-molecules. Thus, development of surface-specific spectroscopic technique operating in the THz range is desired. Indeed, several attempts had been made to study the THz response of surface/interface using optical rectification, but with necessarily large nonlinearity of, e.g., the free carriers in metal[15] or photo-carriers in semiconductors[16]. More recently, THz emission from two-dimensional materials has been reported, particularly, monolayer graphene[17] and transition-metal dichalcogenides[18], thanks to strong enhancement through their electronic resonances. However, in these studies, chemical selectivity through vibrational resonances is not available. More importantly, to probe the surface/interface of complex oxide in general, a versatile surface-specific THz spectroscopic scheme is needed. In this work, we develop a surface-specific terahertz spectroscopic technique using difference-frequency mixing (DFM) process at an interface. The weak terahertz radiation from the interfacial species is detected using the state-of-the-art electro-optic sampling (EOS) technique. Besides complex oxides, our novel terahertz spectroscopic scheme with sub-monolayer sensitivity is expected to benefit the studies of chiral vibrations of biomolecules and collective motions of hydrogen-bonding network at interfaces as well[19, 20].

**Difference-frequency spectroscopy**

Being a second-order nonlinear process, DFM is forbidden in the bulk of centro-symmetric medium under electric dipole approximation, but is allowed at the surface/interface, where translational continuity is necessarily broken. As for THz difference-frequency spectroscopy (THz-DFS), akin to the formalism of SFS, the field spectrum of generated THz pulse can be expressed as[12]:

$$E_{\text{THz}}(\Omega) \propto \chi_{s,\text{eff}}^{(2)}(\Omega) \int d\omega \times E(\omega) E^*(\omega - \Omega) \quad (1)$$

with

$$\chi_{s,\text{eff}}^{(2)}(\Omega) = \chi_s^{(2)}(\Omega) + \chi_B^{(3)}(\Omega) \int E_0(z) e^{i\Delta k z} dz \quad (2)$$

and $E(\omega)$ is the spectrum of incident pump pulse. Here, $\chi_{s,\text{eff}}^{(2)}$ contains both $\chi_s^{(2)}$ from surface structure and $\chi_B^{(3)} \int E_0(z) e^{i\Delta k z} dz$ from surface-field-induced polarization in the depletion layer with $\chi_B^{(3)}$ being the third order nonlinear susceptibility of the bulk and $E_0(z)$ being the depth-dependent static surface field[21]. $\Delta k$ is the phase mismatch along surface normal. When $\Omega$ approaches the resonant frequency of $\chi_s^{(2)}$ or $\chi_B^{(3)}$, the DF signal will be enhanced. In the case that the coherence length $l_c = 1/\Delta k$ is much larger than the thickness of the depletion layer, the second term on the righthand side of Eq. (2) reduces to

$$\chi_B^{(3)}(\Omega) \int E_0(z) e^{i\Delta k z} dz \cong \chi_B^{(3)} \int E_0(z) dz = \chi_B^{(3)} \Phi \quad (3)$$

where $\Phi$ is the surface potential[22]. Thus, $\chi_{s,\text{eff}}^{(2)}$ can be used to characterize both the surface structure and the surface potential, $\Phi$. In the study of a buried interface or multilayer materials, the main advantage of THz-DFS over second harmonic spectroscopy is that the former probes a selected side of target interface with chemical sensitivity, while the latter contains contributions from all interfaces.[23]

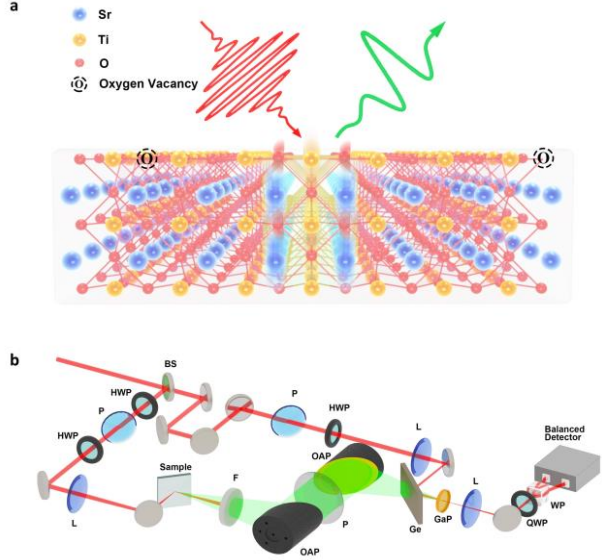

**Fig.1. Schematic of Terahertz difference frequency spectroscopy.** (**a**) Schematic of surface difference frequency generation from STO(001). The incident near-infrared pulse excites TO1 phonon vibration via intra-pulse DFM. The symmetry-breaking of TO1 phonon results from the surface field of STO. (**b**) Sketch of experiment setup. BS, beamsplitter; HWP, half-wave plate; P, linear polarizer; L, lens; F, teflon filter; OAP, off-angle parabolic mirror; QWP, quarter wave-plate; WP, Wollaston prism.

Experimentally, the THz radiation is generated via intra-pulse DFM of a femtosecond pump pulse from a surface/interface (Fig. 1a). Generally, the magnitude of $\chi_s^{(2)}$ for a surface/interface is in the range of $10^{-20} \sim 10^{-22}$ m$^2$/V. Using 1 TW/cm$^2$ pump intensity, the field strength of THz output is estimated to be in the order of $10^0 \sim 10^{-2}$ V/cm. Such a terahertz output can be recorded using the state-of-the-art EOS technique. Recently, we managed to improve the balanced detection in our EOS to a sensitivity[24] of $5\times10^{-8}\ \text{rad}\,\text{Hz}^{-1/2}$, which corresponds to the detection sensitivity of $\chi_s^{(2)}$ reaching $2\times10^{-21}$ m$^2$ V$^{-1}$ Hz$^{-1/2}$ (see Methods). It is sufficient to probe sub-monolayer thick interfacial species with $\chi_s^{(2)} \sim 2\times10^{-22}$ m$^2$/V with acquisition time of 100 s.

**Results**

To examine the validity of THz-DFS for probing the low-frequency resonances at surfaces and interfaces, STO interfaces with or without heterogeneous layer were chosen as the representatives. The schematics in Fig. 1b shows the key components of THz-DFS measurement apparatus (see methods for details). A typical THz-DFS waveform from a pristine undoped STO(001) surface in dry air is presented in Fig. 2a. Here, we used *p*-, *p*- and *p*- (*ppp*) polarization combination for the THz and the femtosecond pump fields. The linear dependence of THz amplitude on pump intensity confirms the signal resulting from the second-order nonlinear process (see Supplementary Material). The Fourier transformed spectrum given in the inset of Fig. 2a shows the detection bandwidth reaches 5 THz that is limited by the EO crystal. After removing the Fresnel factors (see Supplementary Material and Fig. S7), we display the resultant amplitude, imaginary and real parts of $\chi_{s,\,eff}^{(2)}$ spectrum of STO in Fig. 2b, c, and d, respectively. The spectra presented were normalized against a z-cut α-quartz crystal. A single resonant peak is recognized at 2.8 THz with 0.7-THz bandwidth (FWHM) in the $\chi_{s,\,eff}^{(2)}$ spectrum, which can be readily assigned to the TO1 phonon of STO.[25]

Because STO is centro-symmetric in the bulk, the observed DF spectrum must originate from the surface or the surface-electric-field-induced polarization in the depletion layer, corresponding to $\chi_s^{(2)}$ and $\chi_B^{(3)} \int E_0(z) e^{i\Delta kz} dz$ in Eq. (2), respectively[26]. As compared in Fig. 2e, the DF spectrum, $\chi_{s,\,eff}^{(2)}$, of STO shares the same resonant frequency and linewidth with the TO1 phonon of bulk STO[25]. Note that the pure surface mode is expected to differ from those in the bulk in terms of resonant features because of the differences in the structure and local environment in the two cases.[27] Furthermore, we modified the STO surface via deposition of a 20-nm-thick $Al_2O_3$ film. No change in the spectral feature was observed except for increase of the amplitude as shown in Fig. 2e. The above results suggest that the DF spectra, $\chi^{(2)}_{s,\,eff}$, of the STO interfaces are overwhelmed by surface-electric-field-induced polarization, versus the pure surface contribution in *ppp*-polarization. The variance of their amplitudes is attributed to the difference of their surface potential $\Phi$.

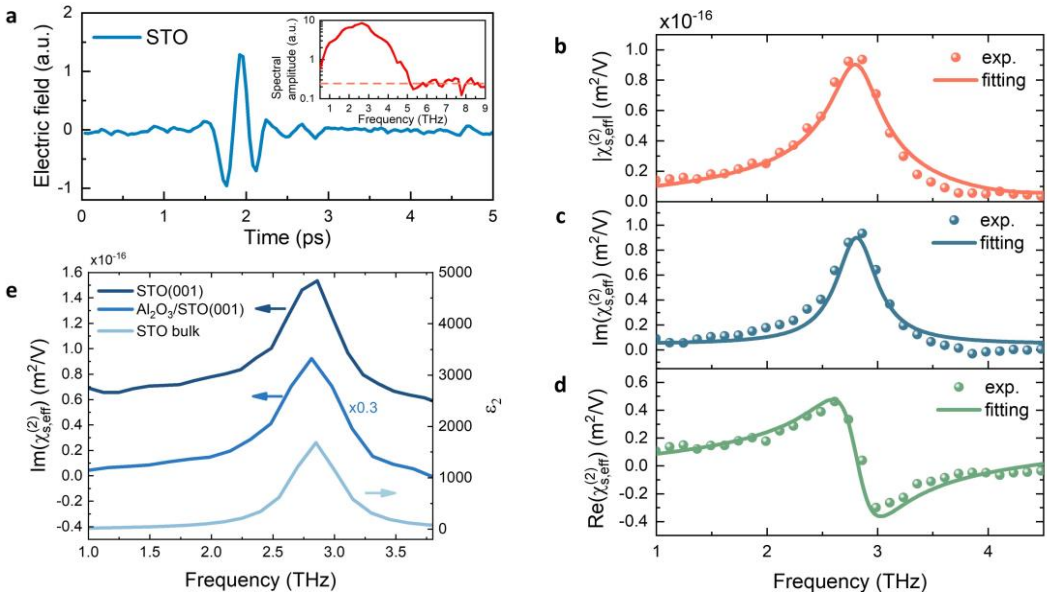

**Fig.2. Terahertz difference frequency spectrum of SrTiO₃.** (**a**) Typical waveform of the emitted terahertz pulse from STO(001) with its Fourier transform in the inset. Dashed line in the inset represents the detection limit. (**b, c** and **d**) Amplitude (**b**), real part (**c**) and imaginary part (**d**) of deduced surface second order susceptibility, after normalization against z-cut quartz reference and removal of Fresnel factor. (**e**) Comparison between Im $\chi^{(2)}_{s,\,eff}$ of bare STO(001) (top), $Al_2O_3$/STO(001) (middle) and the imaginary part of dielectric function ($\varepsilon_2$) of bulk STO (bottom)[25].

To verify that our THz-DFS is sensitive enough to probe a pure surface contribution, i.e., the $\chi_s^{(2)}$ term, we managed to suppress the surface-field induced TO1 phonon contribution by changing the polarization combination from *ppp* to *pss* (*p*-polarized THz field and *s*-polarized pump field). It can be shown that via symmetry argument under *pss*-polarziation, the $\chi_s^{(2)}$ term is non-vanishing at the interface, while the contribution from the polarized phonon in the depletion region is forbiden (see supplementary material and Fig. S2). The interfaces of $Al_2O_3$/STO and $LaAlO_3/SrTiO_3$ (LAO/STO) are known to host two-dimensional electron gas (2DEG).[28, 29] Thanks to the low scattering rate in 2DEG,[30] one expects to observe a Drude-like spectral feature that diverges towards the low frequency according to the hydrodynamic model of free carriers.[31] Indeed, as evidenced in Fig. 3a and b, the nonlinear response, $\chi_s^{(2)}$, from 2DEG is clearly observed in the DF spectra for $Al_2O_3$/STO and 6 unit cell LAO/STO. In contrast, the Drude-like behavior is absent in the DF spectra of 2 unit cell LAO/STO, 50-nm $SiO_2$/STO and air/STO interfaces, in which no 2DEG exists.[28] Note that the nonlinear response of 2DEG is much smaller than the surface field contribution, $\chi_B^{(3)} \int E_0(z) e^{i\Delta k z} \mathrm{d}z$. Through proper selection of polarzition combination, the THZ-DFS is capable of characterizing species at surface/interface using the state-of-the-art EOS detection scheme.

For 2DEG merging at STO interface, the veiled mechanism is still under lively debate, where the interface potential is one of the key parameters that govern the electronic properties. The interface potential of STO can ascribe to surface oxygen vacancy[32, 33] or charge transfer at the heterogeneous interface that causes band bending in the interfacial region.[34] As discussed above, the DF spectrum of TO1 mode in *ppp*- polarization is proportional to the surface/interface potential, $\Phi$ (see Eq. (3)). Thus, once the $\chi^{(3)}_B$ spectrum of STO is calibrated in prior (described in Supplementary Material and Fig. S8, S9), $\Phi$ can be determined *in situ* using THz-DFS. Figure 4a shows *ppp*-DF spectra of gas/STO, SiO$_2$/STO, Al$_2$O$_3$/STO and LAO/STO interfaces. The corresponding surface/interfacial potential is plotted in Fig. 4b. The interface potential of STO with 20 nm Al$_2$O$_3$ overlayer is found to be +0.58 V, which agrees with that determined by X-ray photo-emission spectroscopy (XPS).[35] It confirms the validity of our optical technique for measurement of surface/interface potential. In Fig. 4b, we see no clear correlation between surface/interface potential and the emergence of 2DEG. In particular, the *ppp*-DF spectra for 6 unit cell LAO/STO and 2 unit cell LAO/STO are essentially the same, resulting in the same interface potential of +0.45 V, although the former hosts 2DEG at the interface. It suggests that increasing LAO thickness does not cause obvious change in the interface potential, but may lead to formation of a dipole layer across the interface that creates potential well at the topmost layers of STO for confinement of 2DEG.[36] These results demonstrate THz-DFS as a feasible tool to quantify the surface/interface potential, with unique advantages including chemical selectivity through resonances, functionality for buried interfaces, remote all-optical monitoring, etc.

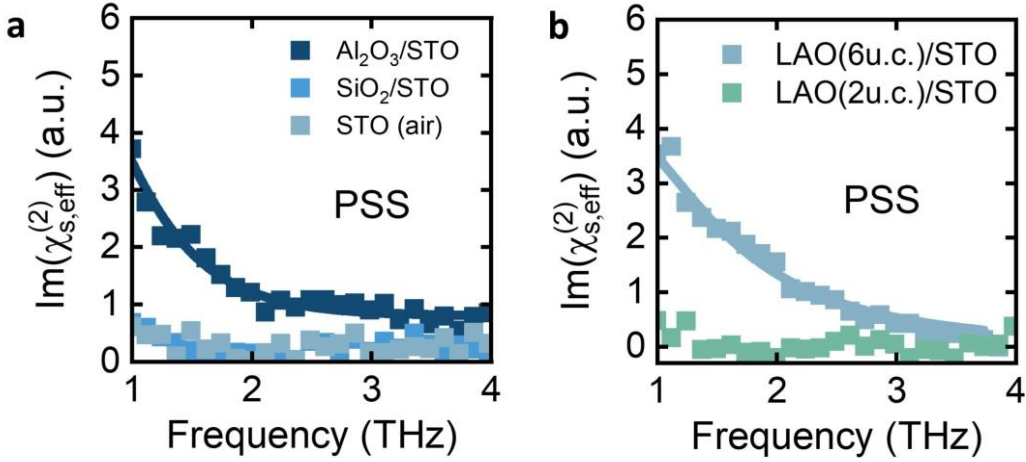

**Fig.3. THz-DFS of 2DEG at STO interfaces.** (**a**) Spectra of Im $\chi^{(2)}_{s,\text{eff}}$ under *pss* polarization combination for bare STO(001) (gray), 20-nm-thick-Al$_2$O$_3$/STO(001) (dark blue), and 50-nm-thick-SiO$_2$/STO(001) (blue). (**b**) Spectra of Im $\chi^{(2)}_{s,\text{eff}}$ under *pss* polarization combination for 2-unit-cell-LAO/STO(001) (green) and 6-unit-cell-LAO/STO(001) (light blue).

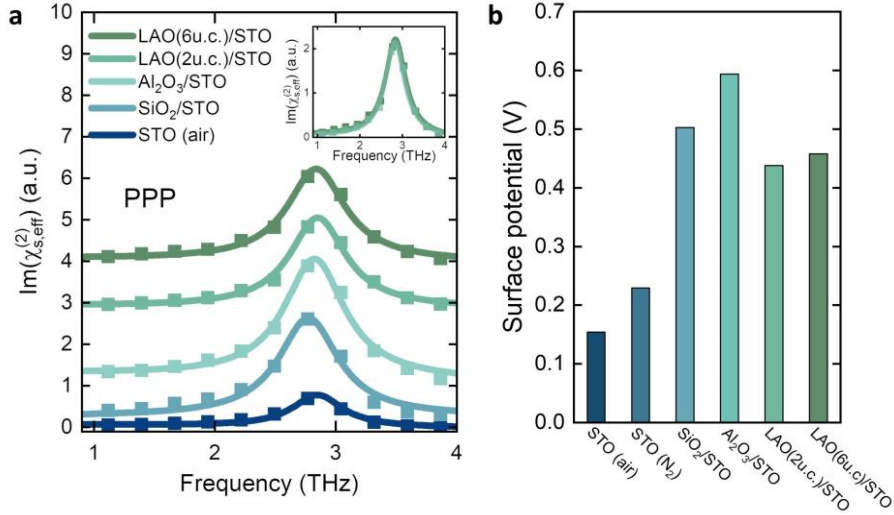

**Fig.4. Surface potential of STO measured by THz-DFS.** (**a**) Spectra of $\text{Im}\,\chi^{(2)}_{s,\,\text{eff}}$ under *ppp* polarization combination, stacked from bottom to top for bare STO, SiO2/STO, Al2O3/STO, 2-unit-cell-LAO/STO, and 6-unit-cell-LAO/STO. Inset, comparison between the spectra of 2-unit-cell-LAO/STO and 6-unit-cell-LAO/STO. (**b**) Summarizing chart of surface/interface potential of STO(001) in various conditions.

**Discussion**

Our work established THz-DFS as a viable surface-specific nonlinear optical spectroscopic method for probing low-frequency resonances on surface or at buried interface. As a demonstration, STO(001) surface and interfaces with different heterogeneous layers were investigated. The 2DEG emerging at the interface and the TO1 phonon polarized in the depletion layer were observed. Furthermore, the sensitivity of THz-DFS also satisfies the detection requirement of sub-monolayer surface species in general using the state-of-the-art EOS technique. In contrast, the well-established sum-frequency spectroscopy found successful applications in probing the elementary vibrations in the mid-IR region, e.g., the chemical and biological materials composed of light elements. Our approach opens up new opportunities for exploring the low-frequency vibrations and emergent species on surfaces or at buried interfaces in various environments.

As an outlook, the detection bandwidth of THz-DFS that we demonstrated here is limited by the EO crystal. Using a high quality organic EO crystal, the detection bandwidth can reach beyond 5 THz. Research in this frequency range is intriguing because collective excitations at the interface of various condensed matter system occur between 5-15 THz, such as superconducting gap, heavy Fermion plasmons, soft mode in ferroelectricity, etc.

**Methods**

*Experimental setup*

The experimental setup of THz-DFS is shown schematically in Fig. 1b. The system is based on a Yb:KGW regenerative amplifier (Light Conversion PHAROS) operating at 100 kHz repetition rate. The amplifier output centered at 1030 nm undergoes a nonlinear spectral broadening stage as described elsewhere.[37] After reflecting on a set of chirped mirrors, the pulse was compressed to 26 femtosecond. A tiny portion was picked out by a beam-splitter to be used as the probe pulse in EOS, while the majority pumps intra-pulse DFM process on the sample surface/interface. The incident pump was 44 μJ in energy per pulse and is focused to 0.5 mm ($1/e^2$ diameter) on sample. An achromatic half wave plate is used to rotate the pump polarization. The emitted terahertz pulse in reflecting direction was collimated and re-focused onto a 0.3-mm-thick GaP(110) crystal by a pair of parabolic mirrors with 200 mm and 100 mm focal length, respectively. Between the two parabolic mirrors, a broadband wire grid polarizer was used as the analyzer for the terahertz radiation. The reflected residual pump was filtered out by a PTFE plate. The whole terahertz beam path was purged with dry air to avoid vapor absorption. Routed through a translational stage, the probe beam was combined collinearly with terahertz beam by a Ge wafer. The polarization change of the probe in EO crystal was measured using the balanced detection scheme consisting of a silicon-photodiode-based balanced detector (Newport Nirvana) and a lock-in amplifier.

*Sensitivity of THz-DFS setup*

The noise spectrum of our THz-DFS measurement setup with integration time over 100 s (10 million pulses) is shown in Fig. S10. Above 1.0 THz the noise level is $2\times10^{-22}$ m$^2$/V, which corresponds to $5\times10^{-8}$ rad$/\sqrt{\text{Hz}}$ in our EOS measurement.[24] Below 1.0 THz, the sensitivity becomes worse towards the low frequency because of the loss by diffraction and the weaker radiation of the oscillating dipoles at lower frequencies. Notice that $\chi^{(2)}_{s,\text{eff}}$ of a self-assembled monolayer is in the order of $1\times10^{-21}$ m$^2$/V. Thus, the sensitivity of our setup is sufficient for the detection of surface-species with sub-monolayer thickness.

**Acknowledgement**

The authors would like to thank Rui Peng and Tong Zhang at Fudan University for supplying sample. C.T. acknowledges the funding support from the National Key Research and Development Program of China (No. 2021YFA1400503 and No. 2021YFA1400202), the National Natural Science Foundation of China (No. 12125403,